\documentclass{aa}
\usepackage{graphics,graphicx,mathrsfs,eufrak}

        \def\Kv{K$_{\rm 2v}$}
        \def\Hv{H$_{\rm 2v}$}
        
\def\vv{{\vec v}}
\def\FF{{\vec F}}
\def\vg{{\vec v_{\rm g}}}
\def\kk{{\vec k}}
\def\xx{{\vec r}}
\def\AA{{\cal A}}

\begin{document}

   \thesaurus{02.08.1; 06.03.1; 02.23.1}     
   \title{Acoustic Waves in a Stratified Atmosphere }
\subtitle{II. Three-Dimensional Hydrodynamics}

   \author{G. Bodo\inst{1} \and W. Kalkofen\inst{2}
   \and S. Massaglia\inst{3} \and P. Rossi\inst{1}
          }

   \offprints{G. Bodo}

   \institute{Osservatorio Astronomico di Torino, Strada dell'Osservatorio
             20, I-10025 Pino Torinese\\
             email: bodo@to.astro.it, rossi@to.astro.it
        \and
        Harvard-Smithsonian Center for Astrophysics, Cambridge, 
        Massachusetts 02138\\
             email: wolf@cfa.harvard.edu
                 \and
        Dipartimento di Fisica Generale dell'Universit\`a, Via Pietro Giuria
        1, I - 10125 Torino\\
              email: massaglia@ph.unito.it                 
             }

   \date{Received:  }

   \maketitle

   \begin{abstract}

We investigate analytically the propagation of linear waves in a 
three-dimensional, nonmagnetic, isothermal atmosphere stratified in 
plane-parallel layers. The motivation is to study oscillations in the
non-magnetic chromosphere and to assess the limitations of 
one-dimensional simulations of the \Kv\ bright point phenomenon.

We consider an impulsively excited acoustic disturbance, emanating from 
a point source, and propagating outward as a spherical acoustic wave 
accompanied by an internal 
gravity wave.  The waves amplify exponentially in the upward direction. 
A significant wave amplitude is therefore
found only in a relatively narrow cone about the 
vertical. 
The amplitude of the wave 
decreases with time. 
Because of the lateral spread, the wave amplitude
decays faster in 2D and 3D simulations than in 1D. The initial pulse,
which travels at the sound speed,  
carries most of the energy injected into the medium.
Subsequent wave crests leave the source region 
at ever-increasing phase speed, but slow to the sound speed as they approach 
the head of the wave. 

Important conclusions from the 3D solution that were not anticipated from the
plane-wave solution are:
\par\noindent
1. The bulk of the energy is emitted in the upward (and downward) direction;
much less goes into the horizontal direction.
\par\noindent
2. The wave profile narrows from the initial pulse through the amplitude
maxima in the wake of the pulse.
\par\noindent
As a consequence of both points, the shock-heated regions in the wake of the
initial pulse would weaken in strength and shrink in size.

\par\noindent
3. The height at which a given wave amplitude is reached spreads outward from
the symmetry axis of the disturbance as the wave propagates upward.
Thus the diameter of the shock-heated region would increase as the acoustic
wave travels upward in the atmosphere.

   \end{abstract}
%

\section{Introduction}

The solar chromosphere exhibits two signatures setting it apart from the 
underlying photosphere: an emission spectrum that characterizes its 
thermodynamic state, and oscillations that characterize its dynamical state. 
Both signatures are present in oscillations in \Hv\ and \Kv\ bright points, 
which arise in the emission peaks on the blue side of the central absorption 
features in the  resonance lines of Ca II, the H and K lines in Fraunhofer's 
nomenclature; these features originate at a height of approximately 1 
Mm above $\tau=1$ in the nonmagnetic chromosphere.

The H and K lines are the strongest lines in the visible chromospheric spectrum
and they show the three-minute oscillations. For these reasons they have 
often been the objects of observational studies (e.g., Liu 1974; 
Cram \& Dam\`e 1983; Lites, Rutten \& Kalkofen 1993), and theoretical 
investigations (e.g., Fleck \& Schmitz 1991; Kalkofen et.\ al 1994, 
hereafter KRBM; Sutmann \& Umschneider 1995; Sutmann, Musielak \& 
Ulmschneider 1998) with the aim of elucidating the excitation mechanism of 
the waves as well as the properties of the chromosphere.

An empirical simulation by Carlsson \& Stein (1997) employing a 
sophisticated radiation-hydrodynamic treatment showed that the 
characteristic features of the spectrum of the H line emerging 
from the chromosphere could be predicted from the observed velocity 
spectrum in the photosphere on the basis of propagating acoustic waves. 
Their model took the photospheric velocity field observed by Lites et al.\ 
(1993) and compared the computed time-dependent emergent H line profile 
with the observed profile. While the model 
gave an emergent H line intensity during the cool phase of the wave, as 
well as an overall temperature structure, that are strongly contradicted 
by the observations (Kalkofen, Ulmschneider \& Avrett 1999), the intricate 
velocity and intensity variations in the line core during the bright phase 
of the wave 
are reproduced to high fidelity; 
only in the timing of the computed \Hv\ intensity variation relative to the 
photospheric motion (and probably also in the absolute intensity) are there 
larger discrepancies from the observations. The otherwise close agreement 
between simulations and observations is surprising since the simulations, 
like all other previous theoretical studies, assume plane acoustic waves. 
The reality of the wave propagation must be more complicated, however, 
suggesting an idealization of cylindrical symmetry. It is therefore 
interesting to investigate the dependence of the results on the geometry of 
the waves and the limitations of the assumed plane symmetry.

We model the excitation of linear hydrodynamic waves and their outward
propagation in a three-dimensional, stratified, isothermal
atmosphere,  assuming that the interaction that gives rise to an outward
traveling wave can be modeled as a pressure pulse at some
reference level in the photosphere. 

Unlike the 1D case, which allowed only acoustic waves to
propagate, the 3D problem admits both acoustic and
internal gravity waves. Our treatment of the hydrodynamic
equations allows the separation of the two modes. We will
focus mainly on the acoustic mode.

Questions to be examined include (1) the lateral spreading of the 
energy as the wave propagates upward,  (2)
the decay of the energy flux from its high value in the initial
pulse to later oscillations in the wake of the pulse and (3) the
increase of the phase velocity from the sound speed at the head
of the wave to infinite phase speeds in the asymptotic limit.

The paper is structured as follows: 
We present the basic hydrodynamic equations in Section 2, describe the 
numerical results from the Fourier solution in Section 3, give the asymptotic 
analysis for late times in Section 4 and summarize the findings in Section 5.
 The appendices contain equations used in the Fourier solution of the 
problem.


\section{The Basic Equations}

The linearized hydrodynamic equations
for adiabatic fluctuations,
 expressing the conservation of mass, 
momentum and  internal energy for a gravitationally stratified, isothermal 
atmosphere can be written in the form:

   \begin{eqnarray}
\nonumber   \frac{ \partial \rho_1} {\partial t} = - \nabla \cdot 
(\rho_0 \vec{v}_1)  \\
\label{syst}
  \rho_0 \frac{ \partial \vec{v}_1} {\partial t}  = - \nabla p_1 - \rho_1 
\vec{g} \\
\nonumber \frac{ \partial p_1} {\partial t} + \vec{v}_1 \cdot \nabla p_0 = 
a^2 \left(  
\frac{ \partial \rho_1} {\partial t} + \vec{v}_1 \cdot \nabla \rho_0 \right) 
\,.
  \end{eqnarray}
where   $p_1$, $\rho_1$ and $\vec{v}_1$ are, respectively, the perturbed  
gas pressure and
density and the velocity; the subscript $0$ denotes the unperturbed 
variables; $\vec{g} = \{0,0,g \} $ is the gravitational 
acceleration, $a = \sqrt{\gamma p_0/\rho_0}$ is the sound speed and $\gamma$
 is the ratio of specific heats.
The equilibrium values of unperturbed variables have the form:
\begin{displaymath}
p_0=p_{00} \exp{(-z/H)}\,,\, \rho_0 = \rho_{00} \exp{(-z/H)} \,,  
\, v_0=0 \,, \, {\rm and}
\end{displaymath}
\begin{displaymath}
   T_0={\rm const} \,,
\end{displaymath}
where $p_{00}$ and $\rho_{00}$ are, respectively, the values of pressure and
density at the reference height $z=0$; and
 $H=a^2/\gamma g$ is the pressure scale height.

In order to simplify the calculations it is convenient
to assume the height dependence (see Lamb 1932)
\begin{equation}
\vec{v}_1 \propto  {\rm e}^{z/2H} \,,
\label{eq:vel} 
\end{equation}
and, as a consequence,
\begin{equation}
p_1  \ , \, \rho_1 \ \propto {\rm e}^{-z/2H} \,. 
\label{eq:pden}
\end{equation}
With these assumptions
and measuring space in units of twice the pressure
scale height, $2 H$, and time in units of
$ \omega_{\rm ac}^{-1}= P_{\rm ac}/2 \pi$,
where $P_{\rm ac} = 4 \pi H/a$ is the cutoff period, 
equations~(\ref{syst}) become
\begin{equation}
\frac{\partial^4 \varphi}{\partial t^4} - (\Delta -1) \frac{\partial^2
\varphi}{\partial t^2} - 4 \frac{\gamma -1}{\gamma^2} \Delta_\perp \varphi=0
\;,
\label{klg}
\end{equation} 
where $\varphi$ is any of the variables (\ref{eq:vel}) or (\ref{eq:pden})
in non-dimensional form, i.e.,
 $\vec{v}=\vec{v}_1/a$,
$p=p_1/p_0$ and $\rho=\rho_1/\rho_0,$
i.e. the `reduced' perturbations, and $\Delta$ and $\Delta_\perp$
are Laplacian operators.
Note that the reduced quantities $p$ and $\rho$
bear a vertical amplification as $ \exp{(z/2H)}$ even though the actual
perturbations $p_1$ and $\rho_1$ decay with $z$.
Note also that Eq.~(\ref{klg}) reduces to the
Klein-Gordon equation in the one-dimensional limit (see, e.g., KRBM).

We perform a  plane-wave analysis assuming that all variables 
are proportional to 
$ \exp [i(\vec k \cdot \vec r - \omega t)]$.
Equation~(\ref{klg}) then yields the 
dispersion relation   
\begin{equation}
\omega^4 - (1+ k^2) \ \omega^2 + 4 {\gamma- 1 \over \gamma^2} k_{\perp}^2 = 0
\,,
\label{disp}
\end{equation}
(Bray \& Loughhead 1974),
where $k^2=k_z^2+k_{\perp}^2$ is in units of $1/2H$
and $\omega$ in units of $\omega_{\rm ac}$. 
Eq.~(\ref{disp}) can be solved to obtain
\begin{equation}
\omega^2={1\over2}{\left[\left(k^2 + 1\right) 
\pm \sqrt{\left(k^2 + 1 \right)^2 - 16 \frac{(\gamma - 1)} {\gamma^2 }
\left(k^2 - k^2_z \right)} \right]} \,
\label{disp1}
\end{equation}
where the plus sign is for
the two roots for {\it acoustic waves} and the minus sign
 for the two roots for {\it internal gravity waves}.

The solution of system~(\ref{syst}),
subject to given initial conditions 
at $t = 0$, can be written in the form of Fourier Integrals:
\begin{eqnarray}
 p=\sum_{j=1,4} \int A_j (\vec{k})\exp[i(\vec{k}\cdot\vec{r}-\omega_jt)] 
d\vec{k} \;\;\;\;\; 
\label{syst7} \\
\rho=\sum_{j=1,4}\int A_j(\vec k)F(\vec k,\omega_j)
      \exp[i(\vec k\cdot\vec r-\omega_jt)] d\vec{k} \;\;\;\;\; \label{syst8} \\
v_z=\sum_{j=1,4}\int A_j(\vec k)G(\vec k,\omega_j)
      \exp[i(\vec k\cdot\vec r-\omega_jt)] d\vec{k} \;\;\;\;\; \label{syst9} \\
\vec k_\perp\cdot\vec v_\perp=\sum_{j=1,4}\int A_j(\vec k)K(\vec k,\omega_j)
      \exp[i(\vec k\cdot\vec r-\omega_jt)]
d\vec{k}\;\;\;\;    \label{syst10}
\end{eqnarray}
The summation goes over four modes corresponding to the four roots of the 
dispersion relation~(\ref{disp}), i.e., both acoustic and internal gravity 
modes travelling in two directions.

In order to obtain the values of $F,G,K$ of Eqs. \ref{syst8}-\ref{syst10},
i.e., the relation between density, velocities 
and pressure in the plane-wave solutions,
 we have Fourier analyzed system (\ref{syst}), derived a set of algebraic 
equations (Appendix A), and
determined the amplitudes $A_j(\vec k)$  by matching the given 
initial conditions at $t = 0$ (Appendix B). As initial conditions 
we have considered
a pressure pulse; the set of conditions becomes: 
 \begin{displaymath}  
p(\vec r, t=0) = \delta p \; \exp{(-{r^2}/{r_0^2})} \\
\end{displaymath}
 \begin{displaymath}  
\rho(\vec r, t=0) = 0 \\
\end{displaymath}
 \begin{displaymath}  
v_z(\vec r, t=0) = 0\\
\end{displaymath}
 \begin{displaymath}  
\vec k_\perp\cdot \vec v_\perp \ (\vec r, t=0) = 0
\end{displaymath}
where $\delta p$ is the amplitude of the initial perturbation
and $r_0$ is the spatial width of the Gaussian. Note that this four conditions 
completely determine the solution since we have fourth order equation. The 
analytical form for
the amplitudes $A_j(\vec k)$ is given in Appendix B.

\section{Analysis of the Results and Asymptotic Solution}

Before we solve these equations it is instructive to consider briefly the
known cases of waves in one-dimensional media.
For a plane wave in a stratified atmosphere the hydrodynamic equations 
admit only acoustic waves, but no internal gravity waves. The solution
for impulsive excitation of a disturbance (see KRBM)
gives an upward traveling pulse that amplifies exponentially with
height; it is followed by a wake that oscillates at the acoustic cutoff
period.
By contrast, a plane acoustic wave due to an impulse in a homogeneous 
medium, for which the atmosphere of our every-day experience is 
a natural approximation,
shows only the signal, but no wake.
It is interesting to investigate the propagation of a disturbance in a
3D medium and to highlight the differences with the 1D cases.
In order to gain some preliminary insight we study the asymptotic behavior
of the solution in 3D.

\subsection{Asymptotic analysis}

 Following Whitham (1974)
we  describe the solution in terms of a slowly varying wavetrain
(i.e., with little variation in a typical wavelength and period), 
writing,
for example, the pressure perturbation as 
\begin{equation}
p(\xx,t) \sim \AA(\xx, t) \exp (i \theta(\xx, t))
\,\, ,
\label{eq:pres}
\end{equation}
where the amplitude $\AA$ and the phase $\theta$ are slowly varying 
functions of position and time. We can then define a 
local wavenumber, $\kk = \nabla \theta$,
and a local frequency, $\omega = - \partial \theta / \partial t$, to obtain for 
the wavenumber $\kk$ the equation of motion 
\begin{equation}
\frac{\partial \kk} {\partial t}  + \left( \vg \cdot \nabla \right) \kk
= 0
\quad,
\label{eq1}
\end{equation}
(see Eq. 11.44 in Whitham, 1974)
where $\vg$ is the corresponding group velocity vector. 
The equation is valid in the limit $t\rightarrow\infty$.

According to Eq.~(\ref{eq1}), the wavenumber $\kk$ is constant along group 
lines, which are defined by the equation,
and each value of $\kk$ propagates with the corresponding constant 
group velocity $\vg(\kk)$. From the initial pulse,
given by the superposition of modes encompassing the whole 
spectrum of wavenumbers, each wavenumber then propagates
with its own, corresponding 
group velocity. A particular wavenumber found at position 
$(x, z)$ at time $t$ can be obtained by solving the pair of equations
\begin{eqnarray}
x = [\vg(k_z,k_\perp)]_\perp t \label{eqn1} \\
z = [\vg(k_z,k_\perp)]_z t \,.
\label{eqn2}
\end{eqnarray}
for the wavenumber components. 
\begin{figure}
\resizebox{\hsize}{!}{\includegraphics{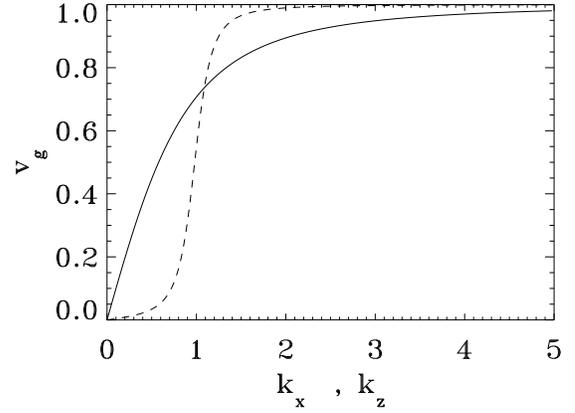}}
\caption{Group velocity vs 
vertical (solid line) and horizontal (dashed line) wavenumbers.}
\label{fig:vgvsk}
\end{figure}
The 
group velocity components in the vertical and horizontal directions are given 
by
 \begin{equation}  
(\vg)_z = \frac{\partial \omega} {\partial k_z} = 
\frac{k_z} {2 \omega} \left[ 1 + 
{ \frac{k^2 + 1} { \sqrt{(k^2+1)^2 - 16 k_\perp^2(\gamma - 1) / \gamma^2 }}} 
\right] \\
\label{vgz}
\end{equation}  
 \begin{equation}  
(\vg)_\perp = \frac{\partial \omega} {\partial k_\perp} = 
\frac{k_\perp} {2 \omega} \left[ 1 + 
{ \frac{k^2 + 1 - 8 (\gamma - 1) / \gamma^2}   
{ \sqrt{(k^2+1)^2 - 16 k_\perp^2(\gamma - 1) / \gamma^2 }}} 
\right]
\label{vgp}
\end{equation}  
where $\omega$ is a root of Eq.~(\ref{disp1}) with the plus sign. 

It is evident, from the dispersion relation (\ref{disp}) or from the
system (\ref{syst7}) - (\ref{syst10}), that the complete 
solution is given by the superposition of acoustic and gravity waves. Here
we confine our analysis to the acoustic waves.

The group
velocity as a function of the moduli of the wavenumbers for the vertical and
horizontal directions is plotted in Fig.~\ref{fig:vgvsk}, which shows 
that the behavior of the two curves 
is different at small and intermediate 
wave\-numbers: in the horizontal 
direction the group velocity remains small up to $ k \sim 0.7 $ and
then increases rapidly to the asymptotic value of unity, and in the vertical
direction the growth begins steeply at $k= 0$ and then continues more 
gradually to the asymptotic value.

\begin{figure}
\resizebox{\hsize}{!}{\includegraphics{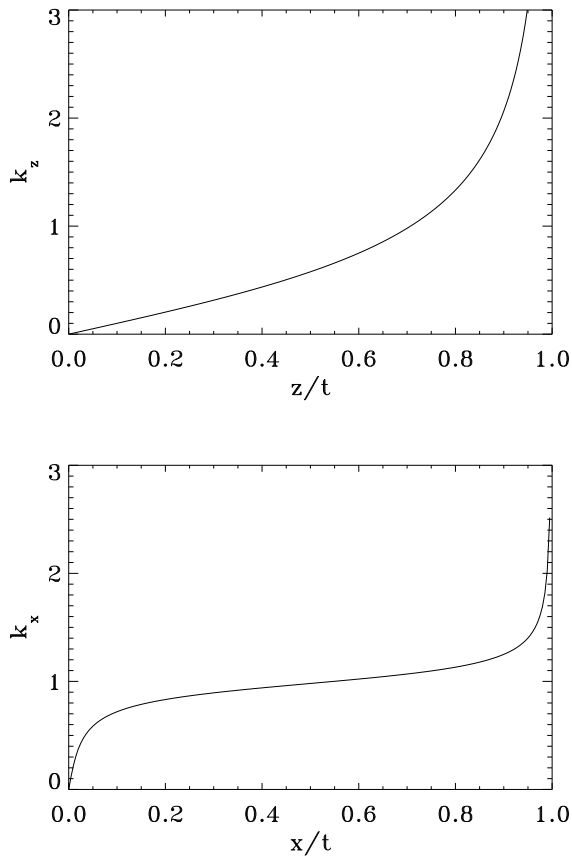}}
\caption{Vertical (upper panel) and
horizontal (lower panel) distributions of wavenumbers.}
\label{fig:vg}
\end{figure}

The evolution of the 
distribution of wavenumbers in space is described by Eqs.~ (\ref{eqn1}) and 
(\ref{eqn2}), with the group velocity $\vg$ given by Eqs.~(\ref{vgz}) and 
(\ref{vgp}).
These equations can be solved for the components $k_z$ and 
$k_\perp$ in the 
vertical and horizontal directions 
as functions of $z/t$ and $x/t$;   the solutions are shown in 
Fig.~\ref{fig:vg}.
The two panels reflect the different behavior,
described above, of the group velocity as a function of wavenumber;
 in both cases, small wavenumbers, which have low group velocity,
 are found close to the origin, while large wavenumbers, which have
group velocity close to the speed of sound, are found near the pulse.
At intermediate distances, the distribution of $k$ values is much broader
 in the vertical than in the horizontal direction. 
This has immediate consequences for the
nature of the solution: since the value of the
wavenumber found at a particular position
implies also the wavelelength of the oscillation, one may expect (see 
Fig.~{\ref{fig:acvsxz}  below) that in the horizontal  direction the 
oscillations behind the front
have almost constant wavelength, while in the vertical direction the
wavelength gradually increases from the head of the wave 
towards the origin.

Combining the phase 
velocity from the dispersion relation
with the distribution of wavenumbers obtained from the evolution equations
(\ref{eqn1}) and (\ref{eqn2}) one can 
follow the propagation of points of constant phase, for example of 
a maximum. 
When a particular maximum is located close to the origin ($x,z=0$)
it is characterized by low values of the  wavenumber, and thus high phase 
velocity.  
As it moves away from the origin, its local 
wavenumber as well as
its frequency increase (in the vertical direction, $\omega^2=k_z^2+1$), while
its propagation velocity decreases  (Fig.~\ref{eq:vel}).
Thus, as a particular maximum travels 
away from the origin and towards the head of the wave it is increasingly 
characterized by high-frequency components
 (which become important for shock formation in the nonlinear regime).
Fig.~{\ref{fig:vpvsxz}} presents this behavior in more detail;
it shows the phase velocity of
three consecutive maxima   as functions of distance from the source
for the vertical (upper panel) and horizontal 
(lower panel) directions.
In the horizontal direction the phase velocity
 is practically equal to the
sound speed nearly throughout the whole region, and for all maxima. 
In the vertical 
direction the behavior is more complicated: 
Except for the head of the wave (the first maximum in Fig.~\ref{fig:vpvsxz}, 
upper panel), 
which propagates at the sound speed, 
the later maxima have high phase velocity 
over increasingly extended height ranges. They therefore travel at increased 
phase velocity (Fig.~\ref{fig:acvsxz}, upper panel; see also KRBM Fig.\ 2), 
especially near the origin (where 
$v_{\rm ph}\rightarrow\infty$ as $z\rightarrow0$), until they approach the 
head of the wave. 
As a consequence of this difference in  behavior between
the two directions, the surfaces of constant phase behind the
wave front travel a longer distance in the vertical than in the horizontal
direction and thus acquire oval shape (see Fig.~\ref{fig:ac} below).

\begin{figure}
\resizebox{\hsize}{!}{\includegraphics{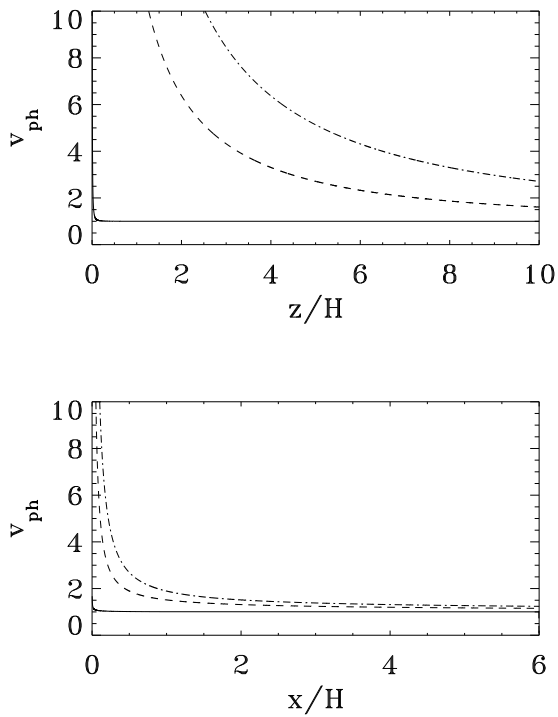}}
\caption{Phase velocity for three consecutive maxima
(solid: first maximum, dashed and dash-dotted: second and third maxima) 
in the vertical (upper panel) and horizontal (lower panel) directions.}
\label{fig:vpvsxz}
\end{figure}

\subsection{Analysis of the results}
We calculate the Fourier integral (7) 
numerically, taking
the symmetry with respect to the $z$-axis
into account, and interpret the results with the aid of the asymptotic 
analysis.
Fig.~\ref{fig:acgrav} 
shows a representative solution for a pressure pulse at $x$$=$$z$$=$$t$$=$$0$,
displaying the reduced pressure distribution,
i.e.,  without the vertical amplification factor $\exp{(z/2H)}$,
in a vertical plane containing the origin, 
at a time $t$ corresponding to about three times the acoustic cutoff
period. As discussed above, the solution is given by the superposition of
acoustic and internal gravity modes. 
The pressure variation due to internal gravity waves,
which appears as radial stripes,
remains confined to the vicinity of the origin since their
group velocity is lower than that of acoustic waves. It is also 
highly anisotropic: internal gravity waves
are excluded from the purely vertical direction,
and in the horizontal direction their group velocity reaches only about
60\% of the sound speed.
Because of their higher
group velocity acoustic waves become more prominent
at greater distance from the origin. They are seen alone in Fig.~\ref{fig:ac}, 
which shows
that the head of the wave forms a spherical surface, with radius equal to 
$a\times t$.

\begin{figure}
\resizebox{4.7in}{!}{\includegraphics[width=\textwidth]{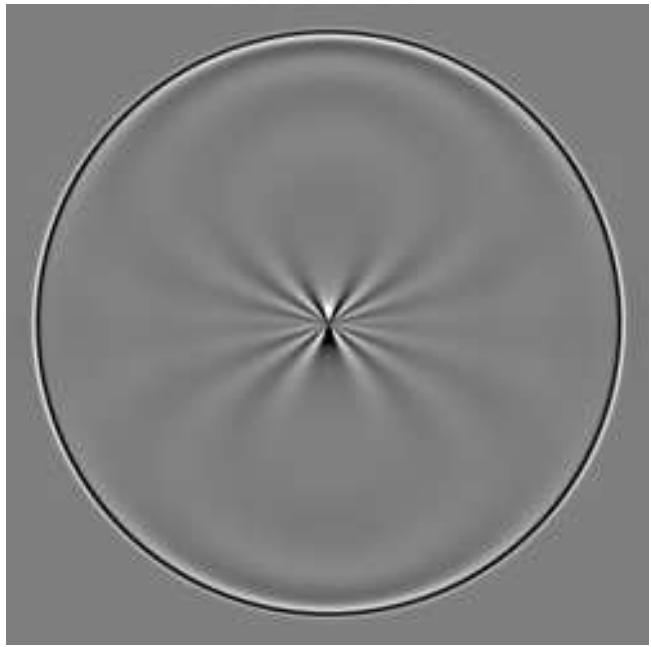}}
\caption
{Acoustic and internal gravity waves at $t=3$ times the acoustic cutoff 
period, in a vertical plane through the origin.
Wave crests and troughs are displayed by bright and dark
shading, respectively. }
\label{fig:acgrav}
\end{figure}

\begin{figure}
\resizebox{4.7in}{!}{\includegraphics{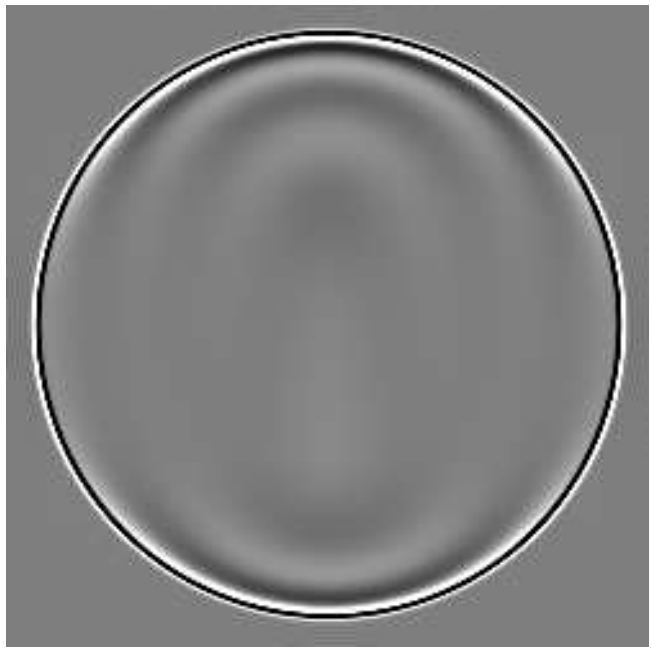}}
\caption
{Acoustic waves  at $t=3$ times the acoustic cutoff period in the
vertical plane.}
\label{fig:ac}
\end{figure}

Recall that in the one-dimensional case 
the solution consists of a wave front followed by a wake 
that oscillates at a period approximately equal to 
the acoustic  period ($\omega=1$). 
In the three-dimensional case this property
is found again in the vertical direction 
(compare Fig.~\ref{fig:acvsxz}, 
upper panel, and KRBM, Fig.\ 2), but now 
also in the horizontal direction, albeit with much reduced amplitude.

The oval shape of the pressure extrema  is related 
to the distribution of $k-$values in space (Fig.~\ref{fig:vg}).
As seen in Fig.~\ref{fig:acvsxz} 
(lower panel), the pressure
amplitude in the horizontal direction has wave crests that are nearly 
equidistant,
consistent with the narrow distribution of $k-$values and, consequently, 
wavelengths, through most of the range of $x-$values. By contrast,
the distance between pressure extrema in the vertical direction increases 
with distance
from the head of the wave due to the corresponding increase in the phase 
velocity, from the sound
speed at the head of the wave to larger values near the origin
(Fig. 2), 
(and to infinitely large values in the asymptotic limit of 
$t\rightarrow\infty$).


\begin{figure}
\resizebox{\hsize}{!}{\includegraphics{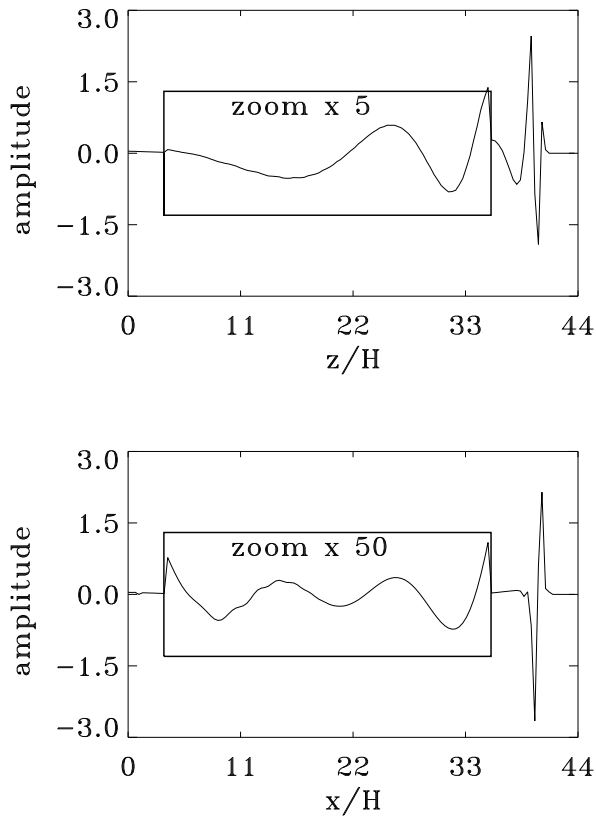}}
\caption
{Vertical (upper panel) and horizontal
(lower panel) cuts of the pressure perturbation
amplitude (in arbitrary units) at $t=3$ times the acoustic cutoff period.}
\label{fig:acvsxz}
\end{figure}

Since the disturbance is due to a point source, the wave starts out as a 
spherical wave. The pulse retains that shape but as can be seen in 
Fig.~\ref{fig:ac}, 
the amplitude of the pulse increases from the vertical direction (both 
positive and negative) towards the horizontal direction.

The three-dimensional solution allows us to investigate
the variation of the perturbation amplitude with
direction relative to the vertical. We can look to this 
variation from two different perspectives: one possibility is
that of considering the variation of the amplitude of a maximum
at a fixed time and the other is that of considering it at a fixed height
(see Fig.~\ref{fig:cartoon}). In an astrophysical context this second
case would
correspond to looking at a fixed optical depth and therefore to
a particular spectroscopic signature of the wave.
In the first case (fixed time), the variation of the amplitude  
has two parts, the intrinsic variation of the reduced function, and the 
variation of the exponential factor, $\exp(z/2H)$, that compensates for the
exponential dependence of the background mass density on height and converts 
the reduced variables into the physical variables. 
But at a fixed height
the distinction between reduced and physical variables is irrelevant.

\begin{figure}
\resizebox{\hsize}{!}{\includegraphics{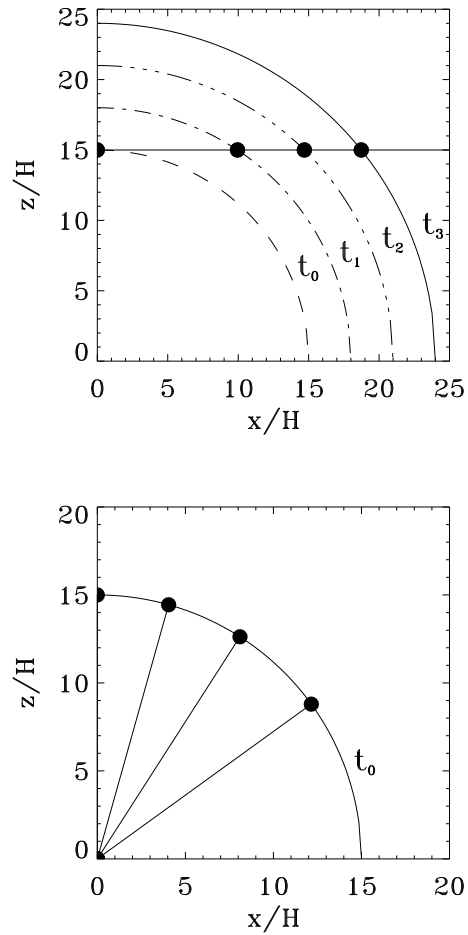}}
\caption
{Upper panel represents the positions of a wave crest at different times
and their intersections with height of $15H$. 
Lower panel shows a wave crest whose apex has reached
the height of $15H$ at time $t_0$.
}
\label{fig:cartoon}
\end{figure}

We consider first the case of fixed height: Fig.~\ref{fig:aczfix} 
shows the variation of the amplitude of the pulse and of the first three
maxima of the wave train as they cross the height of $15H$,
as functions of the angle measured relative to the vertical direction.
Note that, in the actual solar chromosphere, nonlinear effects would already be
present at this height for typical initial perturbations; 
however, our motivation is to study the basic linear behavior of the 
wave propagation. 
Since the various points of the pulse, for example, reach the target height 
of $15H$ with position $x$ and hence angle increasing with time (cf..\ 
Fig.~\ref{fig:cartoon}), the variation of the pressure amplitude reflects 
partly the 
spherical 
decay of the pulse with distance $r$ from the origin
($p\sim r^{-1}$),  and partly the increase of the amplitude with angular
distance from the  vertical, as seen in Fig.~\ref{fig:ac}. The net effect is a
relatively slow  decrease of the pressure amplitude, which reaches a factor
of 2 at  $60^{\circ}$. The decay is much faster for the wave crests in
the wake, reaching a factor of 10 at $60^{\circ}$
for the first and  second maxima, and at $40^{\circ}$ for the third maximum.
The wave profile  evidently narrows considerably with time.

\begin{figure}
\resizebox{\hsize}{!}{\includegraphics{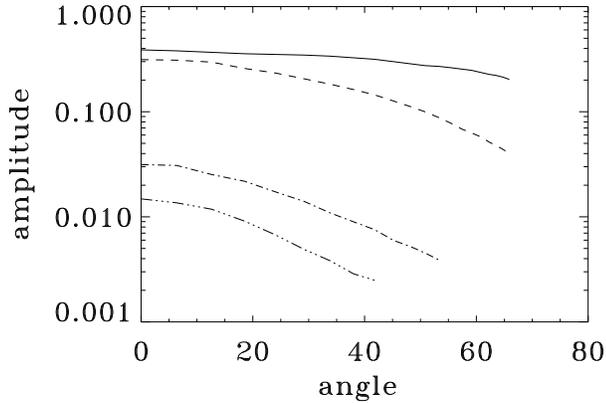}}
\caption
{Angle dependence of the (reduced) pressure amplitude
relative to the vertical direction ($0^{\circ}$). The values 
are for the fixed height of
$15H$ and refer to the pulse (---), and the 1st ($- -$),
2nd ($-\cdot-$) and 3rd ($-\cdot\cdot\cdot-$) maxima of the wake. 
}
\label{fig:aczfix}
\end{figure}

A picture complementary to Fig.~\ref{fig:aczfix}
 is shown in Fig.~\ref{fig:actfix}, 
with the angle variation of the pressure perturbation
at the times when the apices of the pulse and, subsequently, of the
maxima in the wake reach the height of  $15H$.
The upper panel gives the angle dependence of the reduced pressure  
(at the time when the apex reaches $z=15H$), 
and the lower panel that of the physical pressure.
The upper panel confirms the impression from Fig.~\ref{fig:ac} that the 
(reduced) pressure amplitude of the pulse grows with zenith angle. The crests
of the wake weaken with separation from the vertical; they also have
significantly lower amplitude.  The influence of the exponential factor on
the physical pressure is clearly seen in the lower panel, and it results in a
much faster drop of the amplitudes with zenith angle.  This is especially
true for the maxima in the wake for which the oval shape causes a much faster
decay than for the pulse and therefore a narrowing of the wave profile with
the order of the maximum. For the pressure pulse (Fig.~\ref{fig:actfix},
lower panel), the decay in a half-angle of $45^\circ$ is by nearly two orders
of magnitude, and for the maxima in the wake, by nearly three orders of
magnitude.  Note that this decay reflects to a large extent the shape of the
wave crests and is much less dramatic when the points along the crests have
reached the target height, as described above.   \begin{figure}
\resizebox{\hsize}{!}{\includegraphics{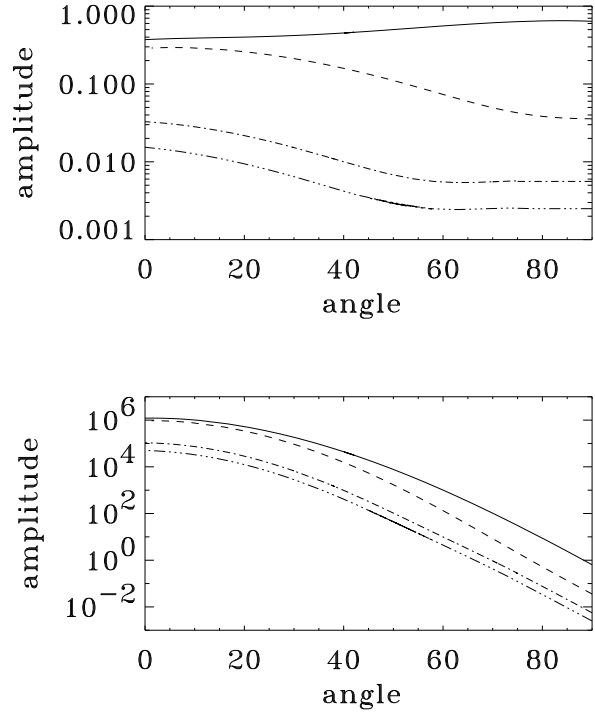}} \caption {Variation
with zenith angle of the pulse (---), and the first three maxima of the wake
($- -, -\cdot-, -\cdot\cdot\cdot-$, resp.) at the instants when the apex for
each feature has reached the target height of $15H$; the upper panel shows
the reduced pressure, the lower panel includes the exponential height factor.
} \label{fig:actfix} \end{figure}  
\section{The energy propagation}

For the system of the linearized equations~(\ref{syst}) 
we can write the conservation  equation
for total energy in the form
\begin{equation}
\frac{\partial E} {\partial t}  + \nabla \cdot \FF = 0
\,,
\label{cons}
\\
\end{equation}
where the total energy density $E$ (in units of $\rho_0  a^2$) is defined as
\begin{equation}
E = \frac{1}{2} \left[ \vv \cdot \vv + \frac{ p^2} {\gamma^2} + \frac{1}{2}
\frac{1}{\gamma - 1} \left(\frac{p}{\gamma} - \rho \right)^2 \right]
\,,
\label{ener1}
\end{equation}
and the energy flux $\FF$ as
\begin{equation}
\FF = p \vv .
\end{equation}
In expression (\ref{ener1}) for $E$ the  terms 
represent, respectively, kinetic 
energy, acoustic potential energy and gravitational 
potential energy. 
For the slowly varying wavetrain
(\ref{eq:pres}) we may consider  energy and energy flux 
averaged over a period, denoted by $\cal{E}$ and $\vec {\cal  F}$,
 where the averaged energy flux may  be written as 
\begin{equation}
\vec {\cal F} =  \vg {\cal E} \,
\label{ener2}
\end{equation}
(Whitham 1974, Eq. 11.69).
This (asymptotically correct) form of the energy flux shows 
the conservation of total energy
in any volume in ordinary space whose boundaries move with the 
group velocity according to Eqs.~(\ref{eqn1}-\ref{eqn2}).
Eq.~(\ref{cons}) can therefore be written as
\begin{equation}
\frac{d{\cal E}}{dt} = -  (\nabla \cdot \vg) \ {\cal E} \,.
\label{cons1}
\end{equation} 
One can also read Eq.~(\ref{cons1}) noting that a positive divergence of the
group lines yields a decay of the energy density.
Now, since ${\cal E} \propto |{\cal A}|^2$ one can derive
the temporal evolution equation for the amplitude $|{\cal A}|^2$,
\begin{equation}
\frac{d|{\cal A}|^2}{dt} = - n \ \frac{|{\cal A}|^2}{t} \,
\label{cons2}
\end{equation} 
(see Whitham 1974, Section 11.6),
where $n$ is the number of dimensions. Eq.~(\ref{cons2}) yields
\begin{equation}
|{\cal A}| \propto t^{-n/2} \,.
\label{cons3}
\end{equation}

Fig.~\ref{fig:acvstime} shows
the temporal decay of the reduced pressure amplitude of the wake after
the pulse has reached  the positions of $z=2 \pi$ on the vertical axis (top
panel) and $y=2\pi$ on the horizontal axis (lower panel). 
The top panel compares
the decay law of $t^{-3/2}$ with the actual behavior
of the wake. The agreement is excellent even though the
analytic result (\ref{cons3}) is valid only for a slowly varying wavetrain. 
Note the (real) time delay
of $\sim  P_{\rm ac}/2$  in the horizontal direction with respect to the 
vertical, a 
consequence of the phase velocity of the wake 
that is smaller in the horizontal direction 
(see Fig.~\ref{fig:vpvsxz}).
Note also that the amplitude of the maxima in the horizontal direction is lower
by a factor of 2 to 3. This follows from the properties of
the energy propagation, as will be discussed below.

\begin{figure}
\resizebox{\hsize}{!}{\includegraphics{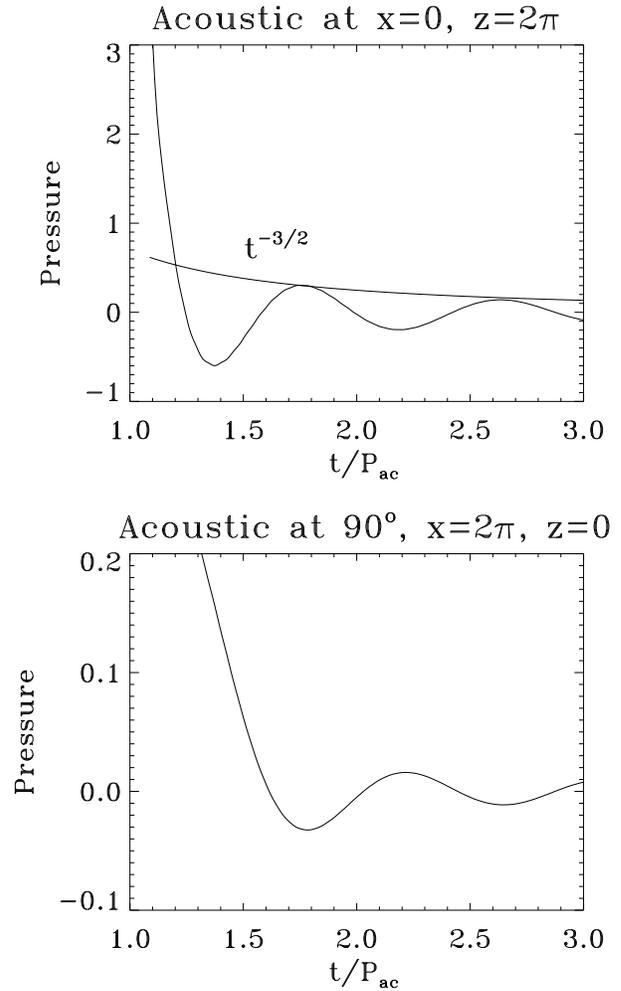}}
\caption{Pressure amplitude of the wake
(in arbitrary units) vs 
time at $x=0$ and $z=2 \pi$ (top panel) and
at $x=2 \pi$ and $z=0$. 
The pulse has been
removed; the amplitude decay of $t^{-3/2}$ is given as a comparison
in the top panel.}
\label{fig:acvstime}
\end{figure}

At $t=0$ the energy is spatially concentrated in the initial pulse, and 
as time elapses, the energy is dispersed into a wavetrain and the wavenumber 
distribution is spread out in  ordinary 
space; both occur with the group velocity, as
described by Eqs. (\ref{eq1}) and (\ref{cons1}). Thus
the energy in any volume
in wavenumber space remains fixed as
the wavenumbers spread in physical space.
Consequently
\begin{displaymath}
\mathfrak{E}(\kk(x,y,z, t)) \ dk_x dk_y dk_z =  E(x,y,z, t) \ dx dy dz \;,
\end{displaymath}
where $E(x,y,z, t)$ is the energy per unit volume 
in ordinary space, estimated
at the point $(x,y,z)$ and time $t$, and
$\mathfrak{E}(\kk(x,y,z, t))$ is the 
energy per unit volume in  the corresponding wave\-num\-ber space, where
$\kk(x,y,z)$ is  the wavenumber 
at position $(x,y,z)$ and time $t$. Now, volumes transform as
\begin{displaymath}
 dk_x dk_y dk_z =  \left| \frac{\partial(x,y,z)}{\partial(k_x,k_y,k_z)} 
\right|
 \ dx dy dz \;,
\end{displaymath}
where 
\begin{displaymath}
{\cal J} =  \left| \frac{\partial(x,y,z)}{\partial(k_x,k_y,k_z)} 
\right| \;
\end{displaymath}
is the Jacobian of the tranformation
between  wavenumbers  and  ordinary space; ${\cal J}$ depends 
on the initial distribution and the expansion or contraction of volumes 
in the mapping from wavenumber space to ordinary space, defined by 
the system (\ref{eqn1}-\ref{eqn2}).
The energy density can therefore be written as
\begin{equation}
E(x,y,z, t) = \frac{\mathfrak{E}(\kk(x,y,z, t))} { \cal J} \;.
\end{equation}

\begin{figure}
\resizebox{\hsize}{!}{\includegraphics{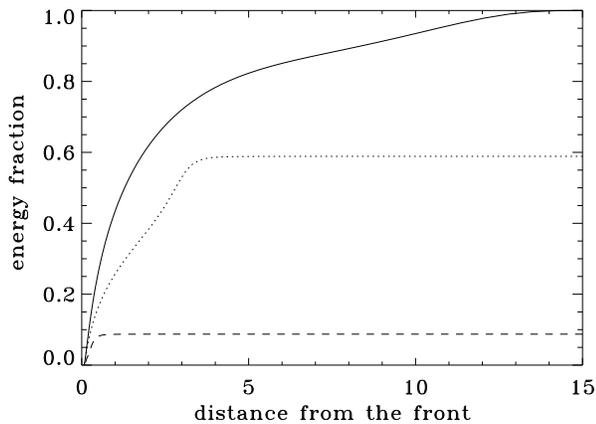}}
\caption{The fractional energy contained inside a cone vs
 distance from the front, for a cone whose axis is 
inclined at 0$^\circ$ (---),  
 45$^\circ$ ($\cdot\cdot\cdot$),  and  90$^\circ$ (- - -)
with respect to the vertical. }
\label{fig:efrac}
\end{figure}

From the above discussion it is clear why  the 
amplitude of the maxima following the pulse is much smaller 
in the horizontal than in the 
vertical direction. The behavior of the group velocity 
and the distribution of $k-$values in space
 (see Figs. \ref{fig:vgvsk} and \ref{fig:vg})
show that in the horizontal direction a narrow  range of $k$ values
near $k=1$ is dispersed over a large portion of physical space, while
the dispersion is much lower in the vertical direction; energy is therefore
much more diluted in the horizontal than in the vertical direction and, 
as a consequence, the maxima have much lower amplitude.

This behavior becomes even more evident from
the fractional energy contained in a cone of infinitesimal
opening angle when its symmetry axis points in different directions.
Fig. \ref{fig:efrac}  shows this fractional energy
at the time when the front
has reached the height of $15 H$, normalized to the total
energy contained in a vertical cone, against 
distance measured from the front of a wave.
Note that in the horizontal direction
the energy rises from zero at the front to the asymptotic value in 
$\sim 0.5 H$. 
This implies that most of the wave energy is contained
in a thin layer, of $\sim 0.5 H$ width, following the pulse. 
But in the
vertical direction about 80\% of the energy is contained within 
a layer with a thickness
of $\sim 5 H$ extending in height from $10H$ to $15H$. 
The wake therefore contains a significant fraction of the energy. 
Approximately 30\% of the energy is 
in the pulse itself, i.e., within a thin layer of $\sim 0.5 H$
at the head of the wave.
Note also the sharp drop of the energy with polar angle in Fig.~\ref{fig:efrac}.
Evidently the bulk of the energy goes into the vertical direction.


\section{Conclusions}
We have carried out a linear analysis of the three-dimen\-sio\-nal 
wave propagation in a
stratified, isothermal atmosphere. The motivation was to study the limitations
of the generally assumed plane waves. Although the analysis was based on
linearized equations and idealized medium our results may suggest the sense
in which 3D and 1D waves differ from one another in the nonlinear regime.

A pulse generated in the `photosphere' by a `point source' propagates 
upward with exponential amplification. 
The source region has a Gaussian pressure perturbation with a 
full half-width of $0.4\times{ H}$, where $H$ is the pressure scale 
height, for a diameter of 40 km. An initial 
perturbation of strength $\delta p/p \approx 0.2$ becomes
nonlinear at a height of $10\times H$. So this strength of the 
source is  near the minimum needed for \Kv\ bright points.

At 1 Mm above the source, 
which corresponds to the height of formation of the H and K lines of Ca II
in the solar chromosphere, the pressure and 
velocity perturbations reach large values over a region with a horizontal 
extent of 1 Mm. This is comparable to the size of Ca  bright points.

The initial pulse is followed by a wake in the upward direction
at the acoustic cutoff period,  approximately three
minutes in the solar atmosphere. There is a wake also in the 
horizontal direction, with the same period but much lower amplitude.

The energy of the wave is concentrated in the vertical direction:
One quarter of the upward-propagating energy is contained within 
a cone with a half-angle of 30$^\circ$ about the vertical axis; that cone 
constitutes only about 13\% 
of the volume of the hemisphere.
The energy is concentrated also in a narrow layer behind the initial pulse: 
When the wave has reached a height of $10H$ above the source, 60\% 
of the energy on the vertical axis is contained within the first two scale 
heights behind the apex. 

The height where a given magnitude of the perturbation is reached increases 
with the size of the region. Assuming that nonlinear conditions are reached 
on the vertical axis at a height of $10H$, a `bright point' with a 
diameter of 1 Mm requires that the wave at the edge travel another $1.5H$, 
which requires an additional 21 s (at a sound speed of 7 km/s), and 
for a diameter of 2 Mm, the additinal travel distance and time are $3H$ and 
45 s. Thus the `bright point' grows from the center outward and upward. 
For the larger diameter and height, correspondingly higher layers in the 
chromosphere would form the spectrum.

The amplitude of the oscillations behind the pulse weakens in strength and 
shrinks in size.
As a consequence, the maximal intensities associated with bright points in 
the wake are weaker and smaller than those in the
initial pulse.

The  maxima in the wake decay with time as 
$v(t)\propto t^{-n/2}$, depending on the geometry.  For a plane wave in a 
one-dimensional medium, $n=1$, the velocity decays as $t^{-1/2}$; it
matches the 
plane wave solution of KRBM; for a line source in a 
two-dimensional medium, the decay is as $t^{-1}$; and for a point source 
in 3D, it is as $v(t)\propto t^{-3/2}$. The energy flux decays as 
$v^2(t)$. Thus the energy contained within the spherical volume of the wave 
decays in step with the increase in volume and hence
the energy within is conserved.

For comparison with observations of chromospheric oscillations
it needs to be borne in mind that the initial atmosphere in our analytic 
solution
is at rest. The numerical simulations by Carlsson \& Stein (1997) compare 
well with observations only when the waves are launched into a disturbed 
atmosphere.

\acknowledgements

This work has been supported by NASA and NSF. GB and PR thank the Smithsonian 
Institution for supporting their visits to CfA.

\appendix

\section{}

System~(\ref{syst}) can be Fourier-analyzed in terms
of $ \vec v_1, p_1, \rho_1 \propto \exp({\pm z/2H}) \ \exp [i(\vec k \cdot \vec 
r - \omega t)]$, with
the $\pm$ sign according to (\ref{eq:vel}, \ref{eq:pden}), to obtain a set
of algebraic equations:
\begin{eqnarray}
i \omega \rho_1+ \frac{\rho_0}{2H} v_{1z} -i \vec{k} \cdot 
\vec{v}_1 = 0 \label{syst4} \\
i \omega \rho_0 \vec{v}_1 - i \vec{k} p_1 + (\frac{p_1}{2H} - 
g \rho_1) \hat z = 0 
\label{syst5} \\ 
i \omega p_1 -i \omega a^2 \rho_1 - (g-\frac{a^2}{2H}) \rho_0 v_{1z} = 
0 \,,.\label{syst6} 
\end{eqnarray}
The linearized variables can be put in a non dimensional 
form, as described in Section
2, and Eqs.~(\ref{syst4}), (\ref{syst5}) and (\ref{syst6}) can be 
solved to obtain 
 $\rho = F p \,, v_z = G p \,,  \vec k_{\perp} \cdot \vec v_{\perp} = K p$, with
$F, G$ and $K$  specifying relations between density, velocities and pressure 
in the plane-wave solutions as:
\begin{eqnarray}
F(\omega,\vec k) = \frac{k^2 + 2ik_z -1} {\gamma \omega^2 - 2 (1 - ik_z)} \\
G(\omega, \vec k) = \frac{i} {\gamma \omega}
{\left[ \frac{\gamma \omega^2 -i \gamma\omega^2 k_z - 2(k_z^2 + k^2)} 
{\gamma \omega^2 -2 (1 -ik_z)}\right]} \\
K(\omega,\vec k) = \frac{k_{\perp}^2} {\gamma \omega} 
\end{eqnarray}

\section{}

The amplitudes $A_j(\vec k)$ for an initial pressure pulse are:
 \begin{displaymath}  
A_1 = A_3 = \delta p \; \exp{(-\frac{k^2}{k_0^2})} \frac{F(\omega_2)} 
{F(\omega_2) - F(\omega_1)} \\  
\end{displaymath}  
 \begin{displaymath}  
A_2 = A_4 = -\delta p \; \exp{(-\frac{k^2}{k_0^2})} \frac{F(\omega_1)}  
{F(\omega_2) - F(\omega_1)} \,,
\end{displaymath}  
where $k_0 = 2/{r_0}$, $\omega_1$ is the solution of dispersion relation 
for acoustic modes and $\omega_2$ is the solution for internal gravity modes.
$A_1$ and $A_3$ are the amplitudes of acoustic modes  while $A_2$ and $A_4$ 
are the amplitude of internal gravity modes.

For an initial velocity pulse  we obtain:

 \begin{displaymath}  
A_1 = - A_3 =  - \delta p \; \exp{(-\frac{k^2} {k_0^2})} \frac{K(\omega_2)} 
{K(\omega_1) G(\omega_2) - K(\omega_2) G(\omega_1) } \\
\end{displaymath}  
 \begin{displaymath}  
A_2 = - A_4 =  - \delta p \; \exp{(-\frac{k^2}{k_0^2})} \frac {K(\omega_1)} 
{K(\omega_1) G(\omega_2) - K(\omega_2) G(\omega_1) }
\end{displaymath}
The solution has then been obtained by performing numerically the
Fourier integrals~(\ref{syst7}-\ref{syst10}).

\end{document}